\documentclass[twocolumn]{emulateapj}
\slugcomment{Submitted to ApJ}
\shorttitle{Quark-Hadron Crossover, QHC19}
\shortauthors{G. Baym et al.}
\newcommand{\Slash}[1]{{\ooalign{\hfil$#1$\hfil\crcr\raise.167ex\hbox{/}}}}

\newcommand{\gtsim}{\protect\raisebox{-0.5ex}{$\:\stackrel{\textstyle >}{\sim}\:$}}

\newcommand \beq{\begin{eqnarray}}
\newcommand \eeq{\end{eqnarray}}

\usepackage[colorlinks=true,linktocpage=true,linkcolor=blue,citecolor=blue]{hyperref}

\begin{document}
\title{New Neutron Star Equation of State with Quark-Hadron Crossover}

\author{Gordon Baym}
\affil{Department of Physics, University of Illinois at Urbana-Champaign, 1110 W. Green Street, Urbana, Illinois 61801, USA\\
RIKEN iTHEMS Program, Wako, Saitama 351-0198, Japan}

\author{Shun Furusawa}
\affil{RIKEN iTHEMS Program, Wako, Saitama 351-0198, Japan}
\affiliation{Department of Physics, Tokyo University of Science, Shinjuku, Tokyo, 162-8601, Japan}

\author{Tetsuo Hatsuda}
\affil{RIKEN iTHEMS Program, RIKEN, Wako, Saitama 351-0198, Japan}

\author{Toru Kojo}
\affil{Key Laboratory of Quark and Lepton Physics (MOE) and Institute of Particle Physics, Central China Normal University, Wuhan 430079, China}

\author{Hajime Togashi}
\affil{Department of Physics, Kyushu Univ., Fukuoka, 819-0395, Japan \\
RIKEN Nishina Center, Wako, Saitama 351-0198, Japan}

\begin{abstract}
We present a much improved equation of state for neutron star matter,  QHC19, with a smooth crossover from the hadronic regime at lower densities to the quark regime at higher densities.   We now use the Togashi et al.~equation of state, a generalization of the Akmal--Pandharipande--Ravenhall equation of state of uniform nuclear matter, in the entire hadronic regime; the Togashi equation of state consistently describes nonuniform as well as uniform matter, and matter at beta equilibrium without the need for an interpolation between pure neutron and symmetric nuclear matter.   We describe the quark matter regime at higher densities with the Nambu--Jona--Lasinio model, now identifying tight constraints on the phenomenological universal vector repulsion between quarks and the pairing interaction between quarks arising from the requirements of thermodynamic stability and causal propagation of sound.   The resultant neutron star properties agree very well with the inferences of the LIGO/Virgo collaboration, from GW170817, of the pressure vs. baryon density, neutron star radii, and tidal deformabilities.  The maximum neutron star mass allowed by QHC19 is 2.35 $M_\odot$, consistent with all neutron star mass determinations.  

\end{abstract}


\section{Introduction}
    It is expected that the cold dense matter in neutron stars is nucleonic (hadronic) at lower densities and becomes a strongly interacting quark phase at high densities.  The equation of state of the matter governs the structure of neutron stars.   In this paper we construct a much improved version, called QHC19, of the equation of state of cold dense matter based on a 
smooth  crossover from the hadronic to quark phase (\cite{Masuda:2013,Kojo:2015,Baym:2018}).   Such crossover equations of state, labeled QHC (for Quark--Hadron Crossover),  are based on a realistic nucleonic matter equation of state at low to medium densities, a strongly interacting quark matter equation of state at high density, and a thermodynamically consistent (and highly constrained) interpolation between the two regimes.  See also \cite{Annala:2019} for a closely related approach to the equation of state.
       
     The new features we take into account are: first, a consistent equation of state for nucleonic matter (\cite{Togashi:2013, Togashi:2017}) from neutron drip in the crust to about 2$n_0$ in the hadronic interior, where $n_0$ is the nuclear saturation density, $\sim$ 0.16 fm$^{-3}$, corresponding to $2.7\times 10^{14}$g cm$^{-3}$); and second we identify and implement strong constraints on the two phenomenological parameters used in the high density quark phase, the universal quark repulsion of coupling strength $g_V$, and the Bardeen--Cooper--Schrieffer (BCS), or diquark pairing strength, $H$.   The resulting QHC19 equation of state agrees, as we show below, strikingly well with the inference of the equation of state by the LIGO/Virgo collaboration from the binary neutron star merger event GW170817 (see Fig.~\ref{matching2} below), and is fully consistent with their radius and tidal deformability determinations (\cite{Ligo-ns:2018a,Ligo-ns:2018b}).  It also allows for a maximum neutron star mass $\simeq$ 2.35 M$_\odot$, consistent with the recent observation of a 2.14 $\pm$ 0.1 M$_\odot$ neutron star  in the pulsar PSR J0740+6620 (\cite{Cromartie:2019}).   In addition, the QHC19 equation of state can be generalized to finite temperature for use in dynamical simulations of neutron star and neutron star--black hole mergers, as well as in other high-energy astrophysical phenomena, including core-collapse supernovae and cooling of proto-neutron stars (\cite{Furusawa:2017b}).     The full details of QHC are posted on the CompOSE archive at \url{https://compose.obspm.fr/eos/140/}.
     
   This paper is organized as follows. In the following section, we first give a brief overview of QHC equations of state, and then present the essence of the Togashi equation of state for hadronic matter in subsec, 2.1,  the Nambu--Jona-Lasinio (NJL) model for quark matter at higher density in subsec, 2.2, and the interpolation region in subsec. 2.3.
In Sec. 3, we discuss the physical ranges of the phenomenological parameters in the NJL model based on the constraints of thermodynamic stability and the sound velocity not exceeding the speed of light.
In Sec. 4, we discuss the structure of neutron stars described by the QHC19 equation of state, and conclude in Sec. 5.
 
\section{Physics of QHC19}

   At very low baryon density the system is characterized by non-relativistic nucleons interacting via nuclear forces, while at high densities one must take quarks seriously as the physical degrees of freedom.   The equation of state in the low density regime can be found by the techniques of nuclear matter theory, while at high densities, owing to the inability to calculate via lattice quantum chromodynamics (QCD) simulations the equation of state at finite baryon density, one must adopt a phenomenological model of high density quark model; we use the Nambu--Jona--Lasinio model in particular.    Furthermore  the strongly interacting intermediate regime, across which the system transitions from nucleonic to quark degrees of freedom, is still not well understood.   While we interpolate between the two regimes, the interpolation turns out to be highly constrained in order that the matter be thermodynamically stable.
   
\subsection{The hadronic regime}

     In the previous version of our equation of state, QHC18 (\url{https://compose.obspm.fr/eos/139/}), we adopted the Akmal, Pandharipande, and Ravenhall [APR] equation of state (\cite{Akmal:1998})  to describe the liquid nuclear matter interior of neutron stars up to a density $2n_0$.   APR is the result of a thorough variational calculation, including realistic two- and three-body nuclear forces, of the ground state energy of pure neutron matter (PNM) and symmetric nuclear matter (SNM) containing equal densities of neutrons and protons.   However, APR has three drawbacks.
          First, to describe matter in beta equilibrium containing
a finite proton fraction,  one generally makes a quadratic interpolate between the PNM and SNM limits in terms of    
$\alpha= 1-2Y_{\rm p}$:    Here $Y_{\rm p}=n_{ {\rm p} }/n_{\rm B} $, with $n_{\rm B} = n_{\rm n} + n_{\rm p}$ the
 total number density of baryons. The second is that descriptions of nonuniform nuclear matter in the crust do not use the same physics as  APR uses in the liquid nuclear matter interior, and thus joining different equations of state in the crust to the liquid interior is inherently inconsistent.  
Finally APR predicts that the sound velocity exceeds the speed of light at $n_{\rm B} \gtsim 5.5n_0$, densities accessible in the core of high mass neutron stars (this drawback also shared with the Togashi equation of state). 
      
     Togashi et al.~(\cite{Togashi:2013, Togashi:2017}) have recently extended the APR approach, to construct a thermodynamically consistent equation of state that overcomes these drawbacks of APR.   The Togashi equation of state, which we now employ in QHC19,  treats the non-uniform matter in the crust and the uniform matter at medium density within the same framework, and at the same time describes all values of the proton fraction $Y_{\rm p}$ without an interpolation between PNM and SNM.      Specifically we use the Togashi equation of state from density\footnote{In QHC18, we did in fact use the Togashi equation of state up to density 0.26 $n_0$, to construct a smooth match to APR extrapolated to beta equilibrium at higher density, despite APR not describing the inhomogeneous matter in the crust.}  4.74 $\times 10^{-10} n_0$ to 2 $n_0$.

    We briefly summarize the key features of the Togashi equation of state in comparison to APR. 
 Both are based on a Hamiltonian including the two-body Argonne V18 potential extracted by fitting two-nucleon experimental scattering data,  as well as the more empirical three-body Urbana IX potential fit in part to light nuclei.   Both APR and Togashi et al. solve the many-body problem by choosing a variational wave function with parameters determined by minimizing the total energy.  The general form of the wave function is that of a free Fermi gas multiplied by factors that build in two-particle correlations dependent on the relative spin, isospin, and orbital angular momentum of the two particles.    Despite differences in detail both the Togashi and APR calculations of the energy of PNM and of SNM are in good agreement (\cite{Kanzawa:2007, Togashi:2013}). 
    
      An illustrative comparison of the Togashi and APR equations of state near the saturation point in SNM
can be made by expanding the energy per baryon in powers of $x=(n-n_0)/3n_0$ and $\alpha=1-2Y_{{\rm p}}$,
following the convention in the CompOSE repository:
\begin{eqnarray}
\frac{\, E \,}{\, A \,} &= & m - E_0 + \frac{\, x^2 \,}{2}  K_0+ {\cal O}(x^3)
 \nonumber \\
&  & +  \alpha^2 \left( J + x L + {\cal O}(x^2) \right) + {\cal O}(\alpha^4).
\label{convention}
\end{eqnarray}
The saturation energy $E_0$ of SNM is 16.1 MeV in Togashi versus. 16.0 MeV in APR, and the incompressibility $K_0$ = 245 vs. 267 MeV in APR.   Near $n_0$, SNM in the Togashi equation of state is slightly softer than in APR, as indicated by the smaller values of 
the symmetry energy $J$ = 29.1 vs. 34.0 MeV in APR, and the slope parameter, $L$ = 38.7 vs. 63.2 MeV in APR.\footnote{The APR numbers cited above are calculated from the fit function in the original APR paper (\cite{Akmal:1998}).}
In neutron stars,  the Togashi equation of state is slightly stiffer at higher densities than APR, as can be seen in the calculation of a  
1.4 $M_\odot$ neutron star containing nuclear matter only (where in the crust in APR we use the Togashi equation of state for $n_{{\rm B}} \le 0.26n_0$).  The radii  are $R$ = 11.6 km for Togashi vs. 11.5 km for APR, and the tidal deformability, $\Lambda$ = 360 vs. 268,      In contrast, QHC19  (set D, Fig.~\ref{fig:cs2_gv_H}) gives $R$ = 11.6 km  and $\Lambda$ = 350 for a 1.4 $M_\odot$ neutron star.
 
  In the neutron star crust, Togashi et al. assume a single species of heavy spherical nuclei forming a Body-Centered Cubic (BCC) lattice surrounded by a gas of nucleons (\cite{Oyamatsu:1993,Shen:2011}), 
and construct the equation of state by the Thomas--Fermi method.   The crust extends to density $\simeq 0.625\  n_0$,  found by comparing the energy per nucleon, for given average $n_{\mathrm{B}}$ and $Y_{\mathrm{p}}$, for inhomogeneous matter with that for homogeneous matter.    As reported in \cite{Kanzawa:2009}, the properties of neutron star crusts calculated with the Togashi equation of state are consistent with those in previous studies (\cite{Baym:1971,Negele:1973,Douchin:2001}). 
 
   The Togashi equation of state outlined above can be generalized to finite $T$ in terms of the Helmholtz free energy, $F(n_{\mathrm{B}},Y_{\mathrm{p}}, T)$, (\cite{Togashi:2017}) thus providing a unified framework to treat both homogenous and inhomogeneous matter with thermodynamic consistency.   The Togashi equation of state is available at \url{http://www.np.phys.waseda.ac.jp/EOS/} and in the CompOSE archive at \url{https://compose.obspm.fr/eos/105/} 
as a table of thermodynamic quantities over a wide range of baryon density, $n_{\rm B}/n_0$, from $4.7 \times 10^{-10}$ to 38 $n_0$, 
$Y_{\rm p}$ from 0 to 0.65; and temperature $T$ from 0 to $\sim 400$ MeV.

\subsection {Quark degrees of freedom}

     In cold dense matter above baryon densities some 5 $n_0$, quarks become the dominant degrees of freedom.    The baryon chemical potential $\mu_{\rm B}$ is in the range $\sim$ 1.5-2 GeV, where the QCD strong coupling constant $\alpha_s$ is of order unity, too large to allow a perturbation theory calculation of the properties of quark matter.    In this strong coupling regime of QCD, one must take into account non-perturbative effects including the generation of constituent quark masses owing to chiral symmetry breaking (strong quark--antiquark pairing; \cite{Hatsuda:1994}), quark--quark pairing leading to color superconductivity (\cite{Alford:2008}), and the mutual interplay of these effects (\cite{Yamamoto:2006}).  (Although QCD is an asymptotic-free gauge theory, the interactions between quarks and gluons become weak only at very high baryon densities (\cite{Collins:1975}), 
well beyond those in neutron stars.)

    As in QHC18, we describe the non-pertubative regime at baryon densities above $\sim 5 n_0$ by the Nambu-Jona Lasinio (NJL) effective model (reviewed in Ref.~\cite{Buballa:2005}), in which gluon degrees of freedom are taken into account only implicitly.   The model contains quarks ($q$) 
    of three flavors -- up, down, and strange ($u,d,s$) -- and of three colors, a flavor-dependent {\em current} mass $m_q$, as well as four-Fermi interactions in the scalar-pseudoscalar channel, the vector channel and the diquark channel, with coupling strengths $G$, $g_V$, and $H$, respectively.  The Lagrangian describing these interactions is discussed in detail in \cite{Baym:2018}  
where the vector repulsion, $g_V$, is needed for quark matter to be able to support heavy neutron stars. 
 In perturbative QCD, $G$, $g_V$ and $H$ are all related to the single gluon exchange process between quarks, while in the non-perturbative regime, they are taken as independent parameters.   The model also includes the (instanton-induced) six-quark interaction with coupling strengths $K$ and $K'$  (\cite{Baym:2018}).   The new physics in QHC19 is that we now provide tight constraints on the parameters $g_V$ and $H$.
    
   The calculation of the ground state energy in the NJL model, a sum of quark and  lepton contributions, is reviewed in \cite{Baym:2018}.  We take the parameters $G$, $K$  as well as the bare quark mass $m_i$ and the momentum cutoff $\Lambda_{\rm NJL}$ in the NJL model to have their standard vacuum values (the HK parameter set in Table I of \cite{Baym:2018}).   While in the standard description of the NJL model, the parameters $g_V$ and $H$ are determined by a Fierz transform of the one-gluon exchange interaction, we treat them as independent here as the equation of state and the subsequent neutron star structure are, as we see below, rather sensitive to their values.   The effect of $K'$  can be absorbed in the variation of $H$ and $g_V$ as far as the equation of state is concerned, so that we take $K'=0$ throughout.

\subsection{Quark--hadron crossover}

   We now discuss the transition from hadronic to quark degrees of freedom, a problem by no means satisfactorily resolved. 
The simplest picture is that of quark--hadron continuity or a quark--hadron crossover, discussed from the points of views
 of quark percolation (\cite{Baym:1979,Satz:1980,Satz:1998}), of QCD symmetry breaking  (\cite{Schafer:1999,Yamamoto:2006}), and of neutron star phenomenology (\cite{Masuda:2013}).   As the baryon density increases in this scenario, matter evolves from hadronic to strongly interacting quark matter with color superconductivity, without a discontinuous jump (or perhaps with a small jump) in the energy density.   This picture is in strong contrast to the more conventional description of the transition from hadronic matter to weakly interacting quark matter via a first order transition (\cite{Chin:1976}; \cite{Alford:2013}).   The quark--hadron crossover allows a relatively stiff equation of state capable of supporting neutron stars of masses greater than 2 $M_\odot$ with a substantial quark core (\cite{Masuda:2013,Kojo:2015}).
   
   Since neither purely hadronic nor purely quark matter descriptions are reliable in the range 2-5 $n_0$, we construct, as in QHC18, the equation of state by a smooth interpolation between hadronic matter at $n_{\rm B} \le 2n_0$ and quark matter at $n_{\rm B} \ge 5 n_0$, expressing the  thermodynamics in terms of the pressure, $P$, as a function of the baryon chemical potential, $\mu_{\rm B}$.   
  (In the Appendix we examine interpolations with different choices of boundaries; with the boundary of the hadronic matter fixed to $2n_0$, and for the quark matter, to 5, 6, and 7 $n_0$.)
   The interpolation procedure,\footnote{We  take a smooth interpolation as a baseline, but one can accommodate a first order phase transition by perturbing the pressure curve with a kink, although too large a first order transition tends to violate either the causality or two-solar mass constraints (\cite{Baym:2018}).  We emphasize that the interpolation uses the information on hadronic and quark matter only as boundary conditions, and does not extrapolate the equations of state of hadronic or quark matter into the interpolation region.  Generally, extrapolated expressions for the thermodynamics, as well for example as transport coefficients, can behave unphysically and should be avoided where possible.}
reviewed in  \cite{Baym:2018},  must satisfy the constraints obeyed by the pressure, that its first derivative, $\partial P/\partial \mu_{\rm B} = n_{\rm B}$, the baryon density, must be positive, and that it must be convex for all $\mu_{\rm B}$, i.e.,  $\partial^2 P/\partial \mu_{\rm B}^2 = \partial n_{\rm B} /\partial \mu_{\rm B} > 0$; in addition the adiabatic sound velocity, $c_s = \sqrt{\partial P/\partial \varepsilon} =  \sqrt{\partial \ln\mu_{\rm B} /\partial \ln n_{\rm B}}$, cannot exceed the speed of light, $c$.    Given the hadronic pressure at $n_{\rm B} \le 2n_0$, these constraints allow only a limited range of quark pressures at $n_{\rm B} \ge 5n_0$, and furthermore tightly constrain the strength of the universal quark repulsion, $g_V$, and the diquark pairing strength, $H$.    Later we further impose the maximum neutron star mass constraint.

\section{Constraints on Parameters}
 
\begin{figure}[h]
\begin{center}
\includegraphics[scale=0.75]{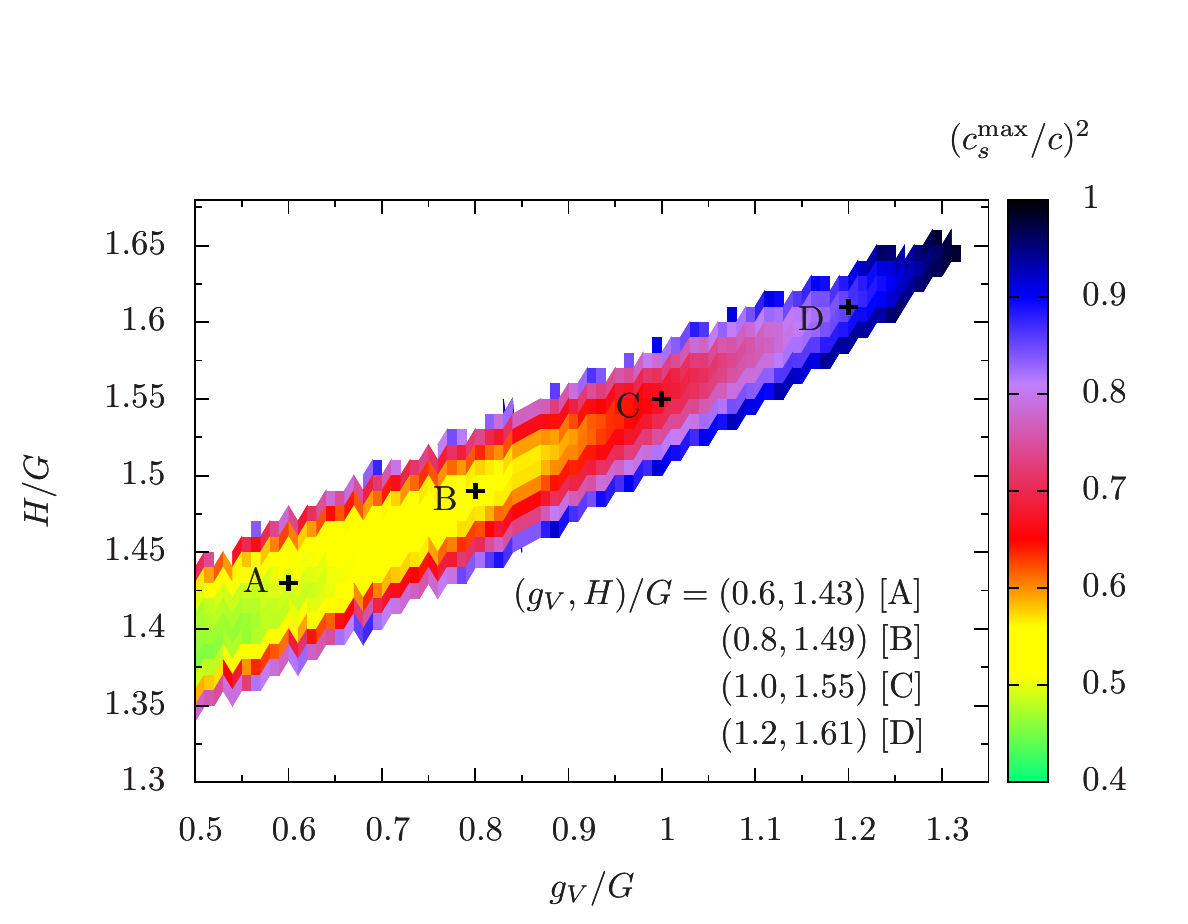}
\end{center}
\caption{The maximal sound velocity squared in the interpolated domain as a function of $H$ and $g_V$. The uncolored region violates either causality or thermodynamic stability, and is excluded.} 
\label{fig:cs2_gv_H}
\end{figure}

   The  quark model parameters that have a particularly important impact on the pressure and hence neutron star structure are  
 $g_V$ and $H$.   We first determine the ranges of these parameters that are consistent with the constraint that the sound velocity, $c_s$, never exceeds the speed of light.
 Figure~\ref{fig:cs2_gv_H} shows $( c_s^{ {\rm max} } /c )^2$ as a function of $g_V$ and $H$; the resolution in the figure is $\Delta g_V =\Delta H = 0.01G$.  The uncolored area violates either causality or thermodynamic stability (i.e., leads to an inflection point in $P(\mu_{\rm B})$ and is unphysical).  As the figure indicates $g_V$ and $H$ are strongly correlated.
 The causality constraint bounds $g_V$ from above by $\simeq 1.3 \,G$, implying that within this NJL description quark matter equations of state for $n_{\rm B} \ge 5n_0$ cannot be too stiff.
 
\begin{figure}[h]
\begin{center}
\includegraphics[scale=0.7]{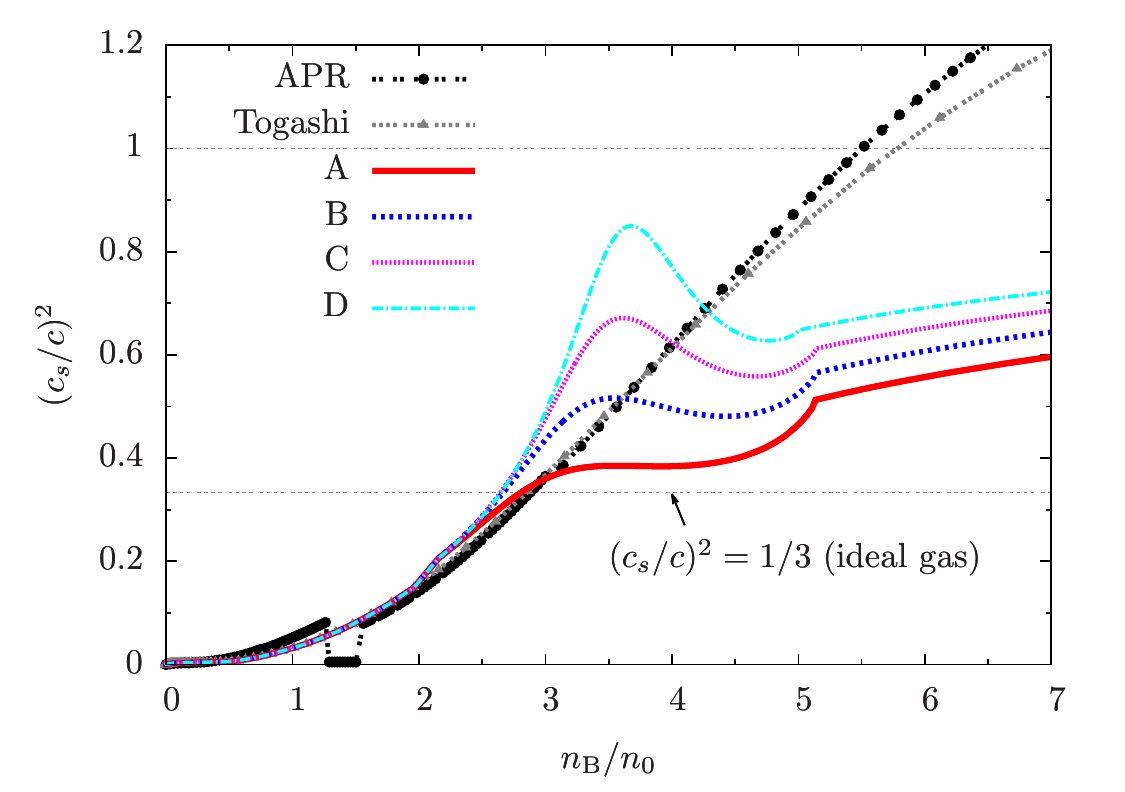}
\end{center}
\caption{Sound velocity squared in units of $c^2$ as functions of $n_{\rm B}/n_0$ for the equations of state of APR, Togashi, and QHC19 for parameter sets A-D, where  $(g_V,H)/G = (0.6, 1.43)$ (set A), $(0.8, 1.49)$ (set B),  $(1.0, 1.55)$ (set C) and  $(1.2, 1.61)$ (set D).}\label{fig:cs2_ABCD}
\end{figure}
 
 As we see in Fig.~\ref{fig:cs2_gv_H}, the maximum sound velocity for given $g_V$  is least in the middle of the allowed region of $H$.
 We focus here on four parameter sets,\footnote{Parameters very close to the boundary of the colored region are unlikely in the sense that for a slight extension to finite temperatures or different lepton fractions the equation of state could easily violate the physical constraints.}  with $H$ chosen to give the least $c_s^{ {\rm max} }$ for given $g_V$: $(g_V,H)/G = (0.6, 1.43)$ (set A), $(0.8, 1.49)$ (set B),  $(1.0, 1.55)$ (set C) and  $(1.2, 1.61)$ (set D), indicated in Fig.~\ref{fig:cs2_gv_H}.


   In Fig.~\ref{fig:cs2_ABCD} we show the density dependence of $c_s^2$ for the sets A--D together with that in APR and Togashi. Replacing a hadronic with a quark equation of state at high density typically reduces the speed of sound. For the case of APR and Togashi, the violation of causality ($c_s>c$) for $n_{\rm B} \gtrsim 5.5n_0$ is cured by this replacement.
   The bump in the QHC19 curves for $n_{\rm B}$ = 3-5 $n_0$ is a consequence of connecting a soft low density equation of state with a stiffer one at high density,  without violating causality. 
(By contrast, the sound velocity at finite temperature and zero baryon density dips between the hadronic and quark regimes; \cite{Asakawa:1997}).)

\begin{figure}[th]
\begin{center}
\includegraphics[scale=0.75]{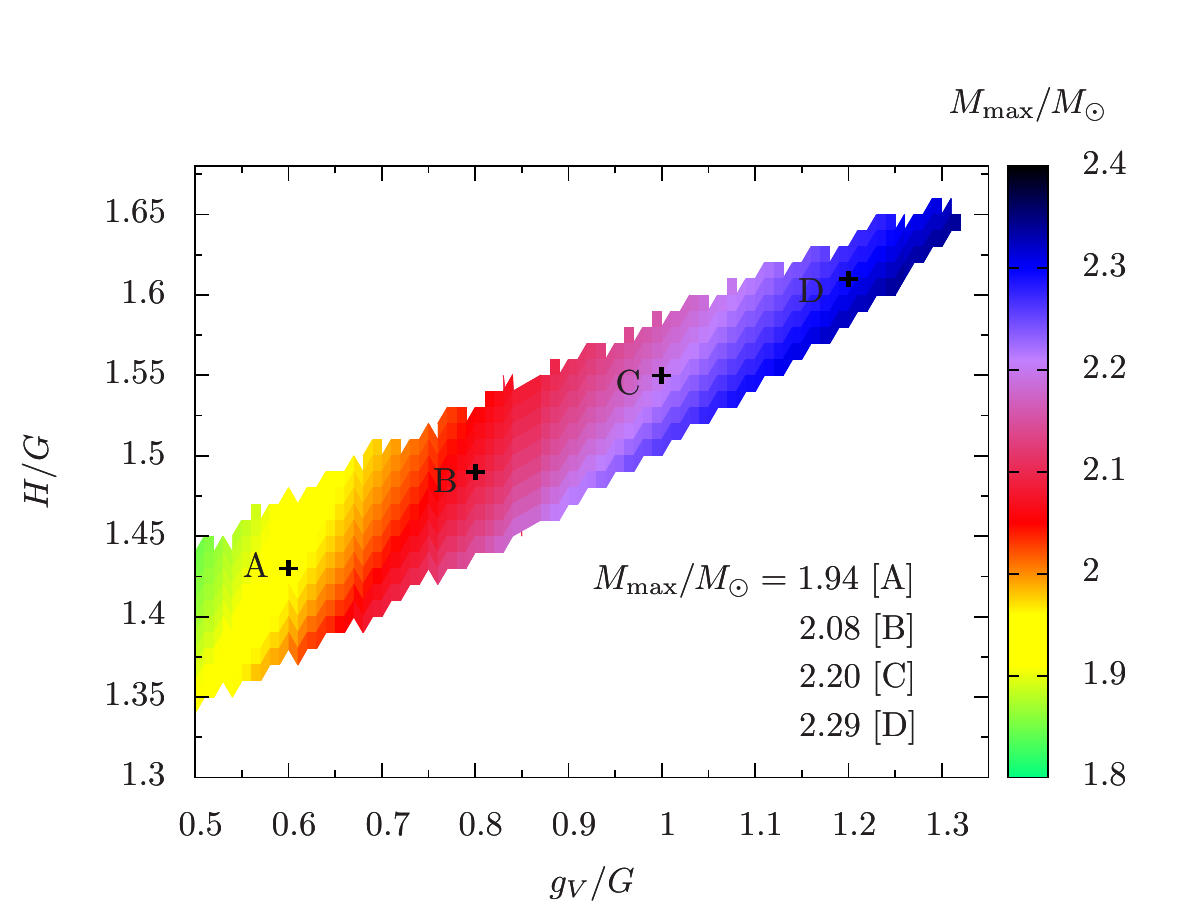}
\end{center}
\caption{The maximal neutron star mass for a given $H$ vs. $g_V$. }
\label{fig:Mmax_gv_H}
\end{figure}

\section{Neutron star structure}

     We now indicate the astrophysical consequences of the QHC19 equation of state for the physical range of $g_V$ and $H$.
Figure~\ref{fig:Mmax_gv_H} shows the maximum neutron star mass, $M_{ {\rm max} }$, for given $g_V$ and  $H$. The absolute maximum allowed mass is $\simeq$ 2.35 $M_{\odot}$ at $(g_V, H)/G \simeq (1.30, 1.65)$,  which is at the causal boundary.   The constraint that the equation of state must be able to support stars of at least two solar masses sets the lower bound, $g_V/G \gtrsim 0.60$-$0.75$, where the detailed value depends on $H$.  The recently determined neutron star of mass 2.17$\pm$ 0.1 $M_\odot$ (\cite{Cromartie:2019})  tightens the constraint to $g_V/G \gtrsim 0.9$.  Beyond the minimal value of $H$, the maximum mass is reduced by increasing $H$ until the equations of state become acausal.

\begin{figure}[th]
\begin{center}
\includegraphics[scale=0.75]{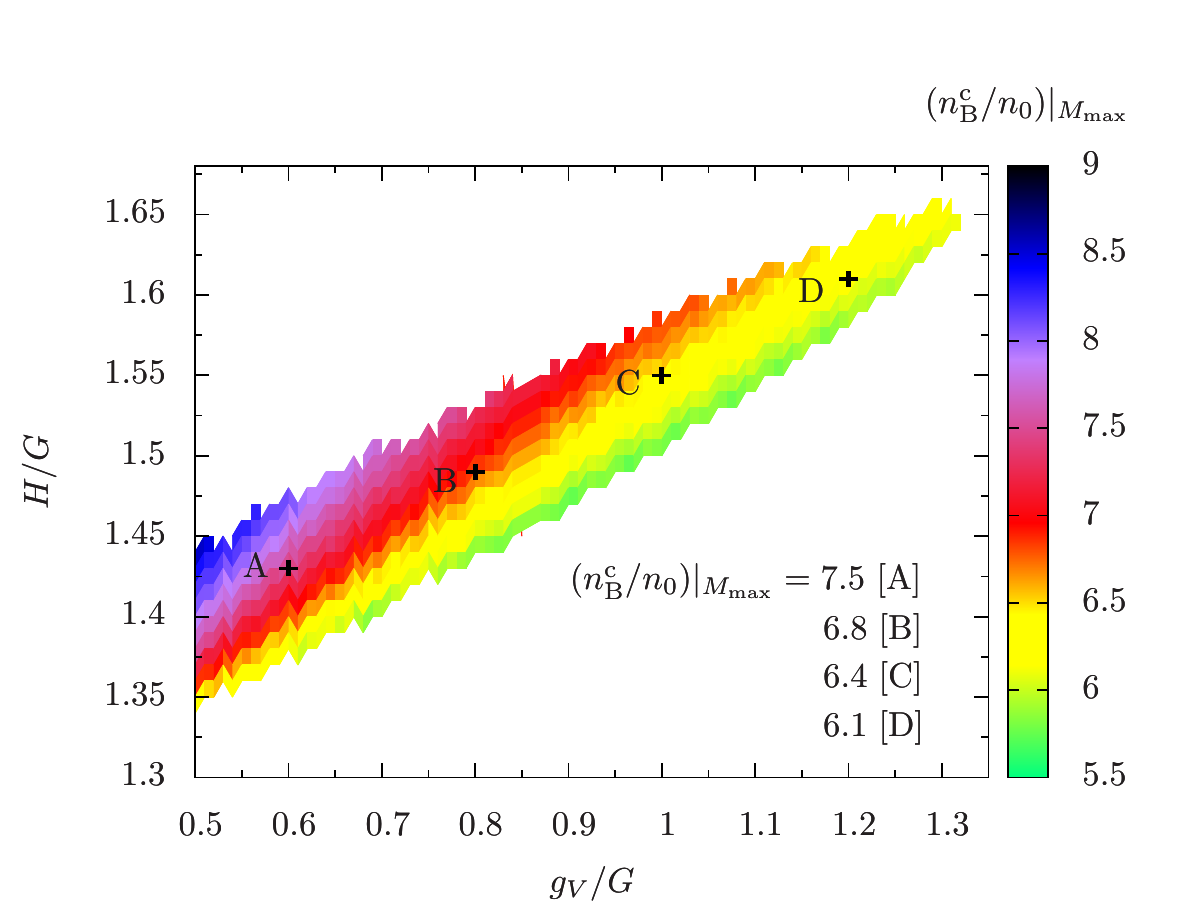}
\end{center}
\caption{ The normalized central baryon density $n_{\rm B}^{ {\rm c} }/n_0$ for a neutron star of the maximum masses for given $H$ vs. $g_V$. }
\label{fig:nb_gv_H}
\end{figure}

   Figure~\ref{fig:nb_gv_H} shows the central baryon density $n_{\rm B}^{ {\rm c} }$ in units of $n_0$ for the 
maximum mass neutron star for each parameter set.  The central density at $M_{ {\rm max} }$ always exceeds $5.5n_0$. 
The central density falls with $g_V$ because the repulsion disfavors large density, while increasing $H$ tends to increase the central density;  although larger $H$ stiffens the equation of state of quark matter, it softens it in the interpolated region, with the net effect that the  central density is larger for larger $H$.

\begin{figure}[th]
\begin{center}
\includegraphics[scale=0.75]{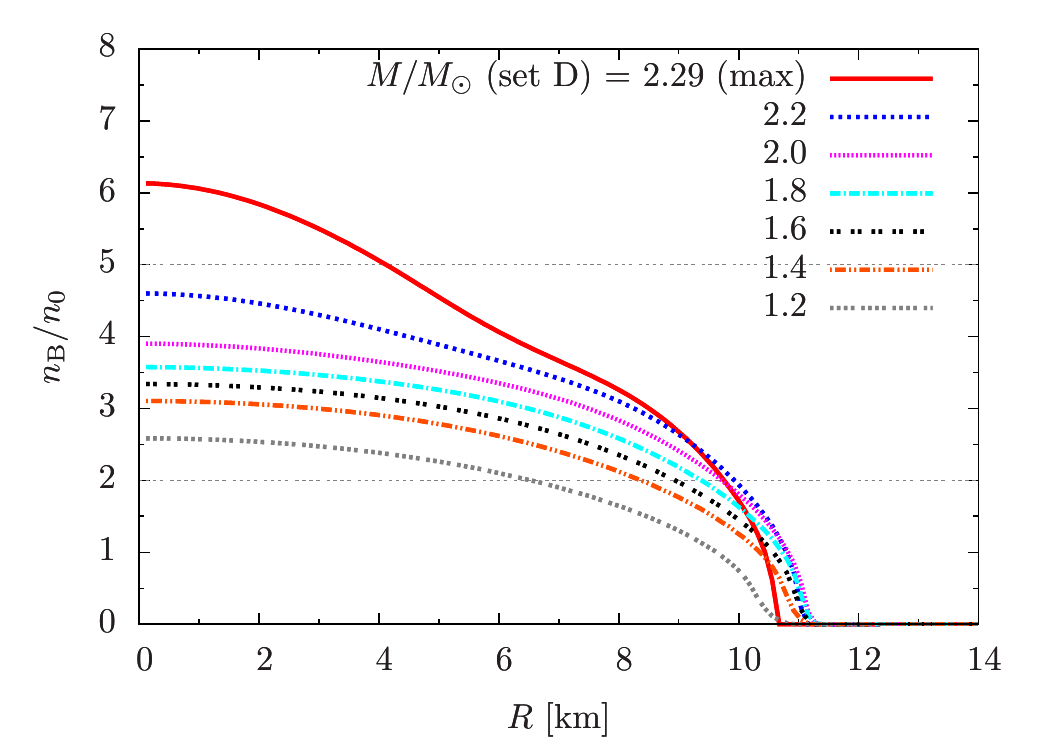}
\end{center}
\caption{The baryon density distributions for QHC19 parameter set D.  The maximal core density is $\simeq 6.1n_0$.} 
\label{fig:nb-R}
\end{figure}

   The baryon density distribution of neutron stars for parameter set D is shown in Fig.~\ref{fig:nb-R}. For $M > 1.2 M_{\odot}$ the core is beyond the pure nuclear regime.  As $M$ increases to $\sim 2.0M_{\odot}$, the core density increases slowly to $\sim 4n_0$. Beyond this region the core density grows more rapidly, reflecting the peak structure of the sound velocity around $n_{\rm B}$ = 3-5 $n_0$ (Fig.~\ref{fig:cs2_ABCD}); after passing the peak the equation of state softens.   

\begin{figure}[th]
\begin{center}
\includegraphics[scale=0.65]{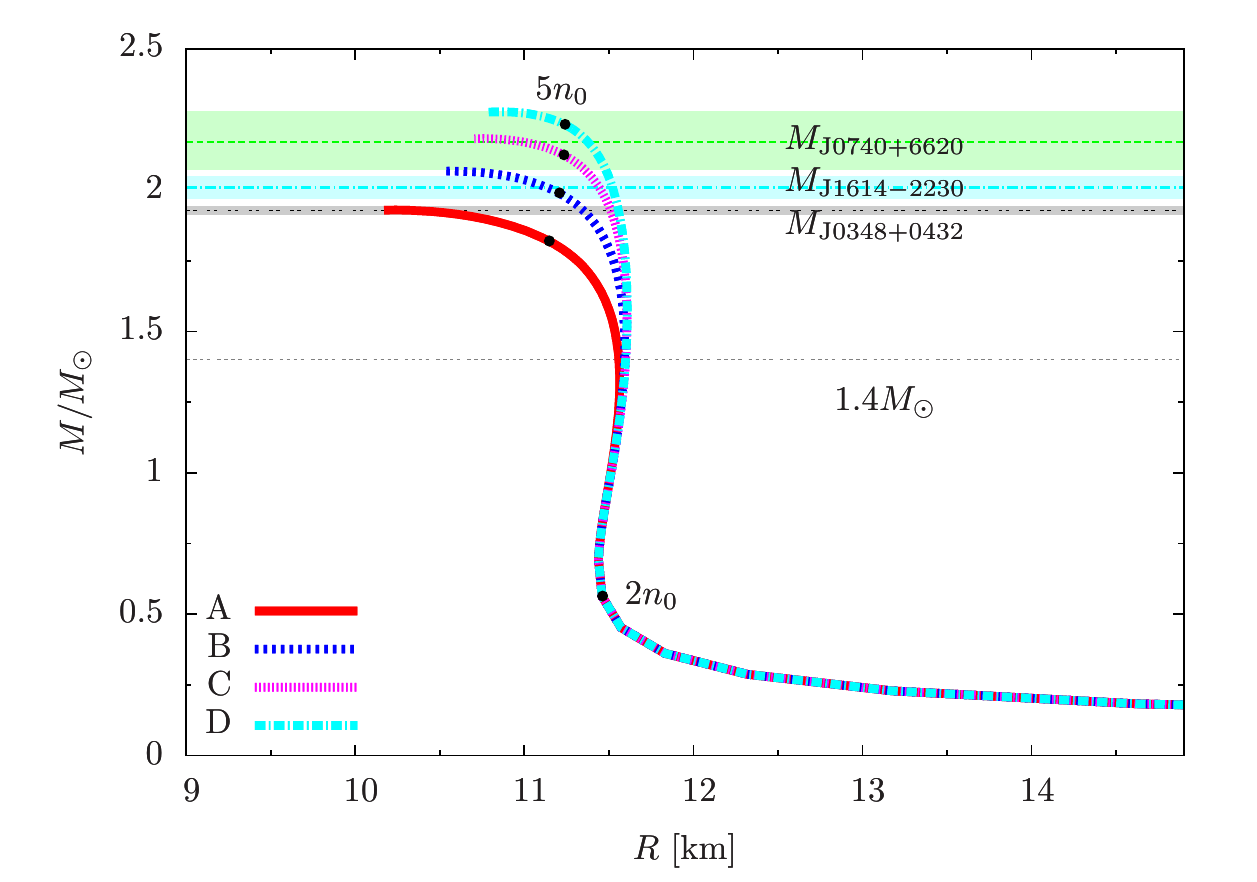}
\end{center}
\caption{The mass-radius relations for the parameter sets A--D.   The baryon density is $2n_0$ at the lower black point on the curves, and $5n_0$ at the upper black points. }
\label{fig:M-R}
\end{figure}

   The neutron star mass-radius relation is shown in Fig.~\ref{fig:M-R} for the parameter sets A--D, together with the masses of PSR J1614--2230 ($M/M_{\odot} = 1.928 \pm 0.017 $), PSR J0348--0432 ($M /M_{\odot}= 2.01\pm 0.04 $), and PSR J0740+6620 ($ M/M_{\odot} = 2.17\pm 0.10 $).   The curves for A--D largely overlap for $M\lesssim 1.4M_{\odot}$ where the corresponding density is $n_{\rm B} \lesssim 2.5$-$3n_0$. The radii at $M=1.4M_\odot$ are $\simeq 11.6$ km, largely due to the relative softness of the Togashi equation of state.  The radii are consistent with the LIGO/Virgo inference $11.9\pm 1.4$ km (\cite{Ligo-ns:2018a}).
   
   We look now at the dimensionless tidal deformability $\Lambda$ of neutron stars (\cite{Hinderer:2008}) calculated with the QHC19 equation of state.     The deformability has a strong impact on the phase evolution of gravitational waves from neutron star mergers, especially at the stage where two neutron stars are not touching but are close enough for the tidal fields to deform each other.  The deformability is strongly correlated with the neutron star compactness, $M/R$; less compact stars deform more easily, and thus have larger $\Lambda$.  The gravitational waveforms in a merger of two stars of masses $M_1$ and $M_2$ are sensitive to the combination,
\beq
\tilde{\Lambda} = \frac{16}{13}\frac{\, (M_1+12M_2)M_1^4\Lambda_1 + (12M_1+M_2)M_2^4\Lambda_2}{ (M_1 + M_2)^5 },  
\eeq
of the masses and $\Lambda_i$ of the individual neutron stars. The individual $\Lambda_i$ can be regarded as functions of the stellar $M_i$, and thus the independent variables in $\tilde{\Lambda}$ are $M_1$ and $M_2$.

\begin{figure}[th]
\begin{center}
\includegraphics[scale=0.85]{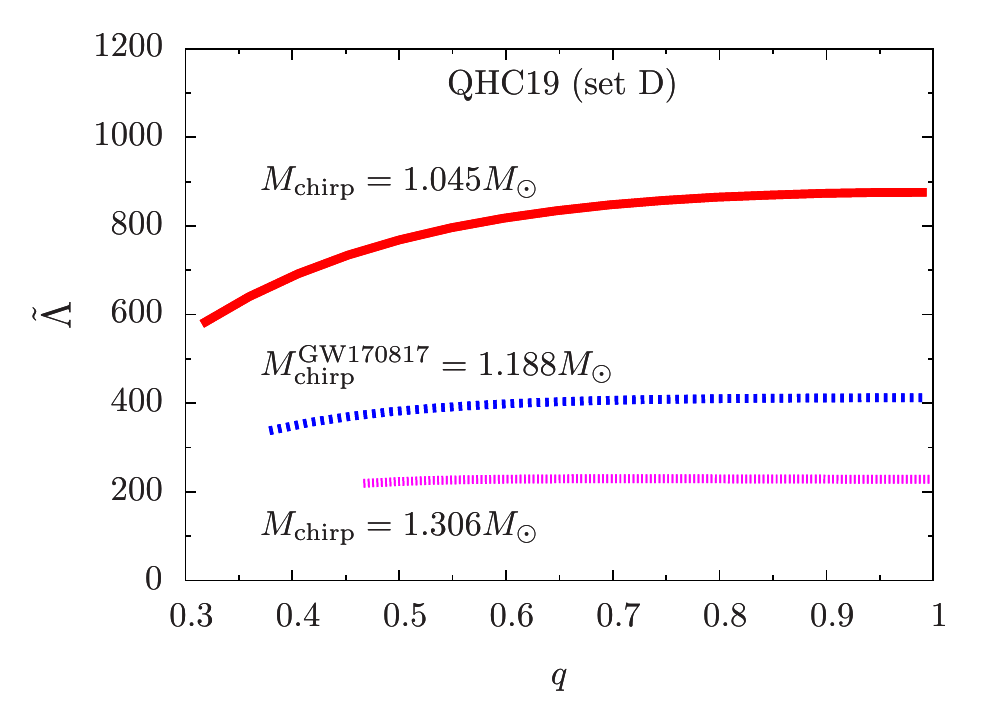}
\end{center}
\caption{The dimensionless tidal deformability $\tilde{\Lambda}$ for given chirp mass, as a function of the mass ratio $q$ calculated with  QHC19 with parameter set D.  }
\label{fig:Lambda}
\end{figure}

    In Fig.~\ref{fig:Lambda} we show $\tilde\Lambda$ calculated with QHC19 for three representative chirp masses,  $M_{\rm chirp}= (M_1 M_2)^{3/5} (M_1+M_2)^{-1/5}$, equal to 1.188 $M_{\odot}$ (corresponding to GW170817), $1.045 M_{\odot}$, and $1.306 M_{\odot}$, as a function of the mass ratio $q=M_1/M_2$.  For equal masses ($q=1.0$) these chirp masses correspond to $M_1=M_2=1.37 M_\odot$, $1.2M_\odot$, and $1.5 M_\odot$.  For the mass ratio $q=0.7$-$1.0$ the parameter sets A-D give very similar curves for $\tilde{\Lambda}$; we display only that for set D.  The calculated $\tilde\Lambda$ for $\mathcal{M}_{ {\rm chirp} } = 1.188 M_{\odot}$ is consistent with the LIGO/Virgo constraints:  for low spin $\tilde\Lambda = 300 ^{+500}_{-190}$ (symmetric interval) 
and $\tilde\Lambda = 300 ^{+420}_{-230}$ (high posterior density interval), and for high spin, $\tilde\Lambda = 0 - 630$.

\begin{figure}[h]
\begin{center}
 \includegraphics[scale=0.71]{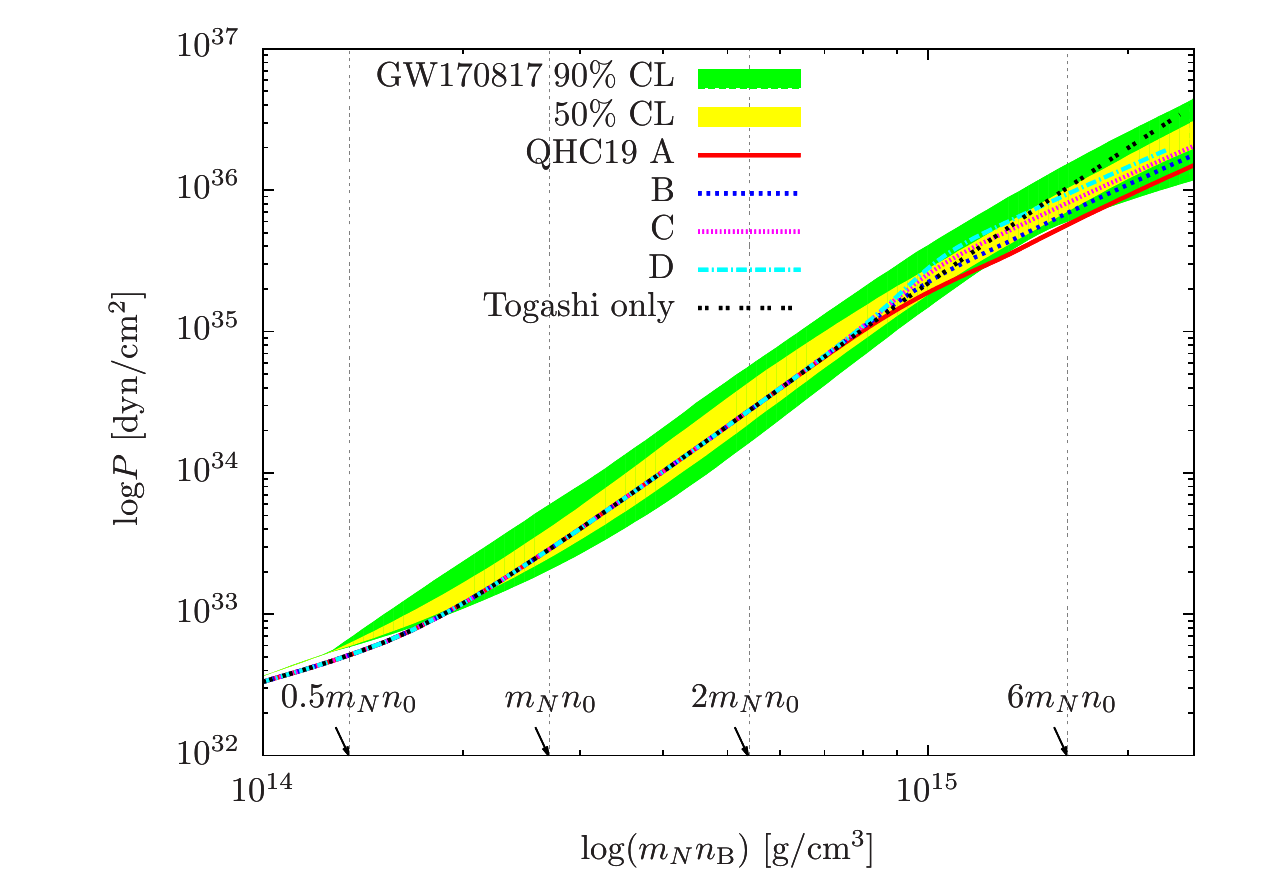}
\end{center}
\caption{Comparison of pressure vs. rest mass density, $m_n n_{\rm B}$ between QHC19 for the parameter sets A-D to the confidence ranges given by LIGO/Virgo (\cite{Ligo-ns:2018a}).  Sets C and D fall within the LIGO/Virgo 50\% band.}
\label{matching2}
\end{figure}

Finally in Fig.\ref{matching2} we compare the QHC19 equation of state, $P$ vs. $m_N n_{\rm B}$, with that recently inferred by the LIGO/Virgo collaboration (\cite{Ligo-ns:2018a}). The latter takes into account the two solar mass constraint (which should now be larger), causality, the inferred tidal deformability, and assumes the nuclear equation of state at low density to be SLy4.    The QHC19 equation of state is remarkably consistent, especially for parameter sets C and D, with the LIGO/Virgo inference,  except at low density due to the difference between the LIGO/Virgo-assumed SLy4 and the Togashi nuclear equation of state.

\section{Summary and Conclusion}

   The new QHC19 equation of state presented here, which includes the physics of strongly interacting quarks at high densities and the  Togashi equation of state in the hadronic regime below 2 $n_0$, well describes neutron stars with masses up to 2.35 $M_\odot$, and is in good agreement with inferences  by LIGO/Virgo, from the GW170817 neutron star merger, of the equation of state as well as radii and tidal deformabilities.  The requirements of thermodynamic stability and causal sound propagation tightly constrain the strength of the universal repulsive interaction and the pairing interaction between quarks.   We await further data on binary neutron star mergers from the present LIGO/Virgo O3 run, as well as from the NICER  X-ray timing observatory on the masses and radii of neutron stars, to further constrain the parameters of the QHC19 equation of state.
   
    There is still room for fundamental improvement, however, of the equation of state, from the crust through to dense quark matter. 
The present description of the nuclei in the crust neglects quantum effects such as pairing, as well as the shell structure of the neutron rich nuclei present.    How the usual shell closures at neutron number 50 and 82 are modified in very neutron rich nuclei, as a consequence of the  nuclear tensor force, remains an open question.  In addition one must take into account nuclear ``pasta" phases in the innermost crust (\cite{RPW,HSY,Watanabe,Caplan:2018}).
    
    Hyperonic degrees of freedom play a role at higher hadronic densities.   Ref.~\cite{Togashi:2016} extends the Togashi equation of state to hyperonic nuclear matter containing $\Lambda$ and $\Sigma^-$ hyperons at zero temperature. 
This study finds that the $\Lambda$ hyperon appears at 0.42 fm$^{-3}$ 
for relatively attractive hyperon interactions. 
This density is well into the interpolation region in this work, where confidence in the validity of hadronic descriptions begins to break down.
Improving our understanding of hadronic interactions, which should be based on quark dynamics at short distance, would allow us to narrow the interpolation domain and to directly discuss the impact of strangeness. In this respect the large uncertainty in hyperon interactions makes it desirable to investigate further effects of hyperons in dense matter with state-of-the-art hyperon--nucleon and hyperon--hyperon interactions from laboratory experiments and from lattice QCD simulations (\cite{Hatsuda:2018}).  

   A major problem in QCD is the lack of a theory of the crossover from hadronic to quark matter, containing both hadronic and quark degrees of freedom.  Efforts in this direction include, e.g., \cite{quarkyonic,yifan}.
   
  In the quark regime, one needs to go beyond the phenomenological NJL model, which does not contain explicit gluons.  
One needs to understand microscopically how the parameters of QHC19 --  the vector coupling $g_V$ and the pairing strength $H$ --  arise from QCD in terms of the strong interaction running coupling $\alpha_s(\mu_{\rm B})$, and how they depend on baryon density.   A step in this direction has been to relate $g_V$ to the QCD quark exchange energy at moderate densities, including effects of non-perturbative chiral and diquark condensates as well as a gluon effective mass (\cite{exchange}).    

  Finally, an equation of state at non-zero temperature is needed for simulations of neutron star--neutron star and neutron star--black hole mergers.   The primary effects of temperature are at lower densities where the Togashi equation of state is already available up to  $T\sim$ 400 MeV.   On the other hand, thermal effects in the quark regime are highly suppressed by the pairing gaps up to $T\sim 100$ MeV.  A finite temperature extension of the QHC19 equation of state  in terms of the Helmholtz free energy, $F(n_{\mathrm{B}},Y_{\mathrm{p}}, T)$, is under investigation and will be reported elsewhere (\cite{hotqhc1}).
 
 \acknowledgments
We are grateful to Vicky Kalogera and collaborators for very useful discussions at the Aspen Center for Physics on equations of state for neutron stars and gravitational wave observations.
The research of author G.B. was supported in part by National Science Foundation Grant No. PHY1714042; 
T.H. was partially supported by the RIKEN iTHEMS program and JSPS Grant-in-Aid for Scientific Research (S), No. 18H05236;
T.K. was supported by NSFC grant 11650110435 and 11875144;
S.F. was supported by JSPS KAKENHI (Grant Nos. JP17H06365 and 19K14723); and
H.T. by JSPS KAKENHI (Nos.~18K13551 and 18H04598).
This work was carried at the Aspen Center for Physics, which is supported by National Science Foundation Grant No. PHY-1607611.

\appendix
 
\begin{figure}[th]
\begin{center}

\includegraphics[scale=0.55]{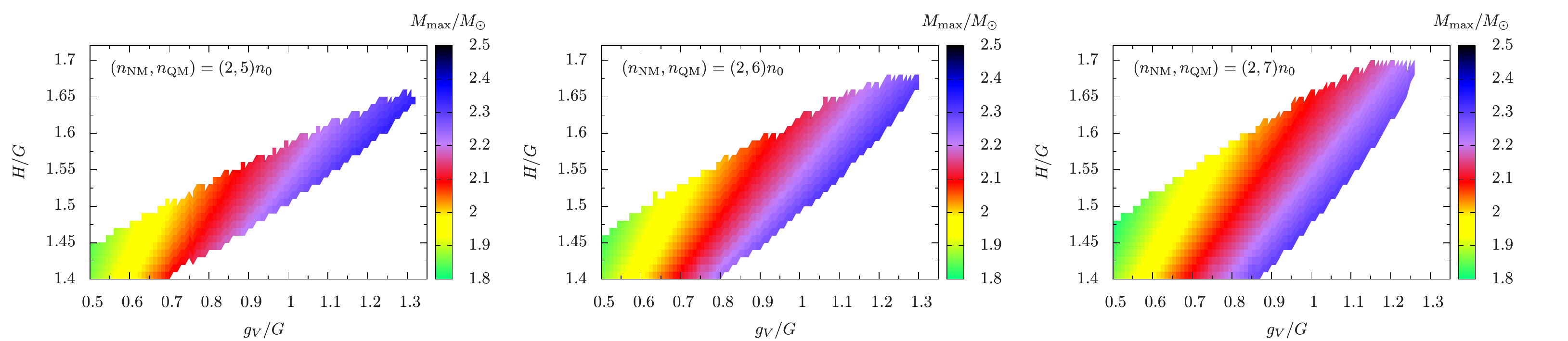}
\end{center}
\caption{The maximal neutron star mass for a given $H$ vs $g_V$. We fix the boundary of the Togashi equation of state to $n_{{\rm NM}}=2n_0$ and the take the quark matter boundary at $n_{{\rm QM}}=5,6,$ and $7n_0$.}
\label{fig:M-R_50-60-70}
\end{figure}

Our interpolation method assumes fixed locations of the boundaries for hadronic and quark matter. As the onset density of quark matter is the major uncertainty, we study upper boundaries $n_{{\rm QM}}=5,6,$ and $7n_0$ while fixing the boundary of the Togashi equation of state at $n_{{\rm NM}}=2n_0$. The results are shown in Fig.~\ref{fig:M-R_50-60-70}. Overall, the extension of the interpolation range relaxes the constraint on the model parameters $g_V$ and $H$ extending their acceptable ranges  by $\sim 10\%$. Meanwhile wider interpolation range also excludes part of the allowed domain for a narrower interpolation range (for example $(g_V,H)/G=(1.3, 1.65)$ is allowed for $n_{{\rm QM}}=5n_0$ but not for $7n_0$). This is simply an artifact due to the insufficiency of our interpolation function with sixth-order polynomials; in order to cover a wider range we need to include higher order polynomials in the interpolation.   We have checked this statement by increasing the number of polynomials whose coefficients, not fixed by the boundary conditions, are varied for some range. These analyses show that the results presented in the main context are good representatives; the change of the quark matter boundary from $5n_0$ to $7n_0$ does not significantly affect our conclusions in the main text. On the other hand, we find that reducing $n_{{\rm QM}}$ from $5n_0$ to $3$-$4n_0$, allowing quark matter at more dilute densities, significantly reduces the acceptable domain for $(g_V,H)$ that is compatible with the constraints from high mass neutron stars.
We have looked at this variation to get a first impression of its effect. A more complete study is desirable but beyond the scope of this paper. 
 
\end{document}